\documentclass[cernpreprint, atlasdraft=false, texlive=2023, USenglish, texmf, orcidlogo]{atlasdoc}
 
\usepackage[subfigure]{atlaspackage}
\usepackage{atlasbiblatex}
 
\usepackage{atlasphysics}

\graphicspath{{logos/}{figures/}}
 
\usepackage{ANA-HDBS-2019-32-PAPER-defs}

\AtlasTitle{Search for Dark Photons in Rare $Z$ Boson Decays with the ATLAS Detector}
 
\AtlasVersion{2.0}
 
\AtlasAbstract{
A search for events with a dark photon produced in association with a dark Higgs boson via rare decays of the Standard Model $Z$ boson is presented, using 139~\mbox{fb\(^{-1}\)}\xspace of $\sqrt{s} = 13$~TeV proton--proton collision data recorded by the ATLAS detector at the Large Hadron Collider.
The dark Higgs boson decays into a pair of dark photons, and at least two of the three dark photons must each decay into a pair of electrons or muons,
resulting in at least two same-flavor opposite-charge lepton pairs in the final state.
The data are found to be consistent with the background prediction, and upper limits are set on
the dark photon's coupling to the dark Higgs boson times the kinetic mixing between the Standard Model photon and the dark photon,
$\ensuremath{\alpha_\mathrm{D}}\xspace{\ensuremath{\varepsilon}\xspace}^2$, in the dark photon mass range of $[5, 40]$~GeV except for the $\Upsilon$ mass window $[8.8, 11.1]$~GeV.
This search explores new parameter space not previously excluded by other experiments.
}

\AtlasRefCode{HDBS-2019-32}
 
\PreprintIdNumber{CERN-EP-2023-107}

\AtlasJournalRef{Phys. Rev. Lett. 131, 251801}
\AtlasDOI{https://doi.org/10.1103/PhysRevLett.131.251801}

\hypersetup{pdftitle={ATLAS document},pdfauthor={The ATLAS Collaboration}}
 
\begin{document}
 
\maketitle

Overwhelming astrophysical evidence~\cite{Bertone:2010zza,Feng:2010gw,Porter:2011nv,Bertone:2017adx} supports the existence of dark matter,
and understanding its nature is one of the most important goals in particle physics.
Dark matter is generally expected to interact very weakly with Standard Model (SM) particles.
This motivates the extension of the SM with hidden or dark sectors~(DS).
One of the simplest examples is an additional \UoneD\ gauge symmetry associated with a gauge boson, the dark photon \dph, that
mediates DS--SM interactions~\cite{Galison:1983pa,Holdom:1985ag}.
In the dark Abelian Higgs scenario, the \UoneD\ symmetry group could be spontaneously broken by a Higgs mechanism through which the dark photon acquires a mass,
adding a dark Higgs boson \hD\ to such models~\cite{Batell:2009yf,Blinov:2017dtk}.

The minimal \dph\ model has three unknown parameters: the mass of the dark photon, \mdph;
the effective coupling of the dark photon to SM particles, \epsl, induced via kinematic mixing with the SM photon;
and the hidden-sector gauge coupling, \alD, which is the coupling of the \dph to DS particles~\cite{Batell:2009yf}.
Dark photons will decay into visible SM particles, either lepton pairs or hadrons, or invisible particles of the DS.
Constraints were placed on visible \dph\ decays, in the parameter space of \mdph and \epsl, by previous beam-dump, fixed-target
and collider experiments~\cite{Batell:2009yf,Marsicano:2018krp,Bjorken:2009mm,KLOE-2:2016ydq,LHCb:2019vmc,CMS-EXO-19-018}.
The dark Abelian Higgs model introduces two additional unknown parameters: the mass of the dark Higgs boson, \mhD, and the mixing between \hD\ and the SM Higgs boson.
The Higgs-strahlung channel, where a dark photon is produced in association with a dark Higgs boson, was also explored at low-energy electron--positron
colliders via $e^+e^- \rightarrow \dph\hD$~\cite{BaBar:2012bkw,Jaegle:2015fme,Belle-II:2022jyy,KLOE-2:2015nli}.
The Higgs-strahlung channel is sensitive to $\alD\epsl^2$, where \alD is also the coupling of the \dph\ to the \hD.
Hence, experimental evidence of a signal in this process would provide information complementary to that from direct searches for \dph.

\begin{figure}[htb]
\centering
\includegraphics[width = 0.33 \textwidth]{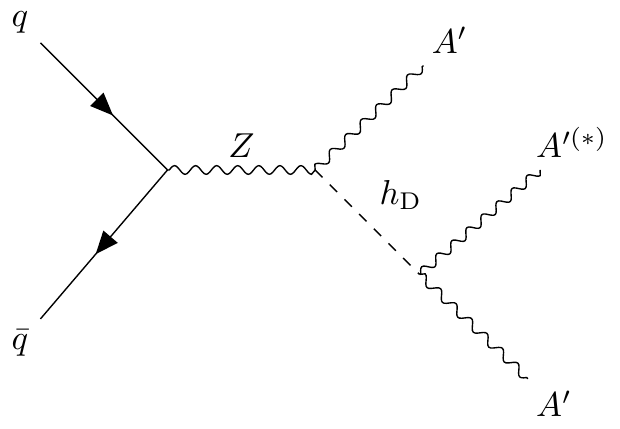}
\caption{Feynman diagram illustrating the signal process $q \bar{q}\rightarrow Z\rightarrow \dph\hD$, $\hD \to \dph \dphs$.}
\label{fig:feynman}
\end{figure}

This Letter presents a search for the dark photon in rare decays of the $Z$ boson, $Z\rightarrow \dph\hD$, with a mass hierarchy of $\mdph + \mhD < \mZ$ and requiring at least two same-flavor opposite-charge lepton pairs in the final state.
For the model considered~\cite{Blinov:2017dtk}, no mixing between the SM and dark Higgs bosons is assumed, the \dph\ is the lightest particle in the DS and invisible DS decays are kinematically forbidden.
When kinematically allowed, the dark Higgs boson can decay into one or two on-shell \dph\ via $\hD \to \dph \dphs$,
as illustrated in Figure~\ref{fig:feynman}, and the \dph\ in turn decays into SM fermions.
The parameter space $\mhD > \mdph$ is explored in this search, giving the process $pp\to Z \to \dph\hD \to \dph\dph\dphs$.
For $\mdph < \mhD < 2\mdph$, the $Z$ boson decays into two on-shell and one off-shell \dph,
and for $\mhD > 2\mdph$, all three \dph are on-shell.
Final states with at least two on-shell \dph\ decaying fully leptonically, $\dph \to \lepPair (\ell = e, \mu)$, are used to search for the \dph.
In this scenario, the kinematic mixing $\epsl$ is small and thus the dark photon has a total decay width narrower than $10^{-3}$~\GeV{},
but $\epsl$ is large enough ($\epsl > 10^{-6}$) to ensure that the dark photon decays promptly~\cite{Curtin:2014cca}.
 
The $\sqrt{s} = 13$~\TeV\ proton--proton~($pp$) collision data used for this analysis were recorded by the ATLAS experiment at the Large Hadron Collider~(LHC) during 2015--2018.
The corresponding integrated luminosity is 139~\ifb~\cite{ATLAS:2022hro} after applying data quality requirements~\cite{DAPR-2018-01}. A combination of single-lepton and multi-lepton triggers~\cite{TRIG-2018-01,TRIG-2018-05} is used.
The ATLAS experiment at the LHC is a multipurpose particle detector
with a forward--backward symmetric cylindrical geometry and a near \(4\pi\) coverage in
solid angle~\cite{PERF-2007-01,ATLAS-TDR-2010-19,PIX-2018-001}\footnote{ATLAS uses a right-handed coordinate system with its origin at the nominal interaction point (IP) in the center of the detector and the $z$-axis along the beam pipe.The $x$-axis points from the IP to the center of the LHC ring, and the $y$-axis points
upward. Cylindrical coordinates ($r$, $\phi$) are used in the transverse plane, $\phi$ being the azimuthal angle around the $z$-axis.
The pseudorapidity is defined in terms of the polar angle $\theta$ as $\eta = -\ln{\tan{\theta /2}}$.
Angular distance is measured in units of \(\Delta R \equiv \sqrt{(\Delta\eta)^{2} + (\Delta\phi)^{2}}\).}.
An extensive software suite~\cite{ATL-SOFT-PUB-2021-001} is used in data simulation, in the reconstruction and analysis of real
and simulated data, in detector operations, and in the trigger and data acquisition systems of the experiment.

Monte Carlo~(MC) simulated signal samples were generated~\cite{Alwall:2014hca},
with matrix elements~(ME) calculated at leading order~(LO) in perturbative QCD and with the \NNPDF[3.0nlo]~\cite{Ball:2014uwa} parton distribution function~(PDF) set.
The events were interfaced to \PYTHIA[8.230]~\cite{Sjostrand:2014zea} to model the parton shower, hadronization, and underlying event, with parameter values set according to the A14 parton-shower tune~\cite{ATL-PHYS-PUB-2014-021} and using the \NNPDF[2.3lo]~\cite{Ball:2012cx} set of PDFs.
Benchmark signal samples were generated with $\alD=0.1$ and $\epsl=10^{-3}$, in the mass ranges $5~\GeV <\mdph < 40~\GeV$ and $20~\GeV < \mhD < 70~\GeV$,
with a mass step of 1~\GeV and 10~\GeV, respectively.
The contribution from $\dph \to \tau^{+}\tau^{-}$ is found to be negligible and thus not included in MC signal samples.
The dominant SM background process, $q\bar{q} \rightarrow 4\ell$, was simulated with the \SHERPA[2.2.2] event generator~\cite{Bothmann:2019yzt}.
Matrix elements were calculated at next-to-leading-order~(NLO) accuracy in QCD for up to one additional parton and at LO accuracy for two and three additional parton emissions.
The ME calculations were matched and merged with the \SHERPA parton shower based on Catani–Seymour dipole factorization~\cite{Gleisberg:2008fv,Schumann:2007mg},
using the MEPS@NLO prescription~\cite{Hoeche:2011fd,Hoeche:2012yf,Catani:2001cc,Hoeche:2009rj}.
An alternative $q\bar{q} \rightarrow 4\ell$ sample for estimating the theory modeling uncertainty was generated at NLO accuracy in QCD using \POWHEGBOX[v2]~\cite{Alioli:2010xd,Melia:2011tj,Nason:2013ydw},
interfaced to \PYTHIA[8.186]~\cite{Sjostrand:2007gs} for the modeling of the parton shower, hadronization, and underlying event, with parameter values set according to the AZNLO tune~\cite{STDM-2012-23}.
The \CT[10] PDF set~\cite{Lai_2010} was used for the hard-scattering processes, whereas the \CTEQ[6L1] PDF set~\cite{Pumplin:2002vw} was used for the parton shower.
The real higher-order electroweak contribution to $4\ell$ production in association with two jets (which includes vector-boson scattering,
but excludes processes involving the Higgs boson) was not included in the sample discussed above but was simulated
separately with the \SHERPA[2.2.2] generator.
\SHERPA[2.2.2] was also used for the $gg\rightarrow ZZ$ process, with LO precision for zero- and one-jet final states, where a constant NLO/LO correction factor of 1.7~\cite{Caola_2015} is applied to account for NLO effects on the cross-section.
The resonant $H\rightarrow ZZ^{*} \rightarrow 4\ell$ process was generated independently to provide the highest possible precision. The dominant gluon--gluon fusion~\cite{Hamilton_2013} and vector-boson fusion (VBF)~\cite{Nason_2010} processes were modeled with \POWHEGBOX[v2]. The gluon--gluon fusion sample used \POWHEG-NNLOPS~\cite{Hamilton_2013,Hamilton:2015nsa,deFlorian:2016spz,Anastasiou:2016cez} to achieve inclusive NNLO QCD precision.
Four or more prompt leptons can also be produced by a number of triboson processes ($VVV$, including $ZWW$, $ZZW$ and $ZZZ$) and by $Z$ bosons produced in association with a $t\bar{t}$ pair ($t\bar{t}Z$). Samples for these $VVV$ and $t\bar{t}Z$ processes were simulated with \SHERPA[2.2.2] and \SHERPA[2.2.0], respectively.
 
Except for the signal, all samples were produced with a detailed simulation of the ATLAS detector~\cite{SOFT-2010-01} based on \GEANT~\cite{Agostinelli:2002hh}, to produce predictions that can be compared with the data.
The signal samples were produced through a simplified simulation of the ATLAS detector~\cite{SOFT-2010-01}.
Furthermore, simulated inelastic minimum-bias events were overlaid to model additional $pp$ collisions in the same and neighboring bunch crossings (\pileup)~\cite{SIMU-2020-01}.
Simulated events were reweighted to match the \pileup conditions in the data. All simulated events were processed using the same reconstruction algorithms as used for data.

Events are required to have a collision vertex associated with at least two charged-particle tracks, each with a
transverse momentum \(\pt\) > 0.5~\GeV{}. The vertex with the highest sum of the squared transverse momenta of the associated tracks is referred to as the primary vertex.
Muon candidates within the range $|\eta|<2.5$ are reconstructed by combining the inner detector~(ID) and muon spectrometer information~\cite{PERF-2015-10}.
In the region $2.5<|\eta|<2.7$, muons can also be identified by tracks of the muon spectrometer alone.
In the region \(|\eta| < 0.1\), muons are identified by an ID track with \(\pt\) > 15~\GeV{} associated with a compatible calorimeter energy deposit.
Muons are required to have \(\pt\) > 3~\GeV{} and \(|\eta| < 2.7\), and satisfy the \enquote{loose} identification criterion~\cite{PERF-2015-10}.
Electrons are reconstructed from energy deposits in the electromagnetic calorimeter matched to a track in the ID~\cite{EGAM-2018-01}.
Candidate electrons must have \(\pt\) > 4.5~\GeV{} and \(|\eta| < 2.47\), and satisfy the \enquote{loose} identification criteria~\cite{EGAM-2018-01}.
All electrons and muons must be isolated and satisfy the \enquote{Loose} and \enquote{PflowLoose\_VarRad} isolation criteria~\cite{EGAM-2018-01,MUON-2018-03}, respectively.
Furthermore, electrons~(muons) are required to have associated tracks satisfying $|d_{0}|/\sigma_{d_{0}} < 5~(3)$ and $|z_{0} \sin(\theta)| < 0.5$~mm, where $d_{0}$ is the transverse impact parameter relative to the beam line, $\sigma_{d_{0}} $ is its uncertainty, and $z_{0}$ is the longitudinal impact parameter relative to the primary vertex.
 
Jets are reconstructed with the anti-$k_t$ algorithm~\cite{ATL-PHYS-PUB-2018-012,Fastjet} with a radius parameter of $R = 0.4$.
The jet-clustering input objects are based on particle flow~\cite{PERF-2015-09} in the ID and the calorimeter.
Jets are required to have $\pT>30$~\GeV{}~and $|\eta|<4.5$.
A jet-vertex tagger~\cite{PERF-2014-03} is applied to jets with $\pT<60$~\GeV\ and $|\eta|<2.4$ to preferentially suppress jets that originated from \pileup.
An overlap-removal procedure detailed in Ref.~\cite{HIGG-2018-28}
is applied to the selected leptons and jets, to avoid ambiguities in the event selection and in the energy measurement of the physics objects.

Candidate events are selected by requiring at least two same-flavor and opposite-charge~(SFOC) lepton pairs.
The four-lepton invariant mass must satisfy $m_{4\ell} < m_{Z} - 5~\GeV{}$ to suppress the SM $pp\to 4\ell$ background.
If more than one lepton quadruplet is selected in an event, the one with the smallest lepton-pair mass difference $|\mllpm - m_{\ell^{'+}\ell^{'-}}|$,
where \mllpm\ and $m_{\ell^{'+}\ell^{'-}}$ are the invariant masses of the two SFOC lepton pairs in the quadruplet, is selected.
The lepton pair with the higher~(lower) invariant mass is denoted by \mllone~(\mlltwo).
To ensure that both SFOC lepton pairs from a signal event originate from an on-shell \dph decay and to reduce the mispairing effect,
the dilepton masses must satisfy $\mlltwo \addslash \mllone > 0.85$.
All the same-~(different-)flavored leptons are required to have an angular separation of $\Delta R > 0.1~(0.2)$.
The two SFOC lepton pairs (and the two pairs with the alternative opposite-charge pairing, in the case of 4$e$ and 4$\mu$ final states) within a quadruplet are
required to have a dilepton mass $\mllpm > 5$~GeV.
Events with $(m_{\Upsilon(1S)}-0.70$ \GeV{}$) < \mllpm < (m_{\Upsilon(3S)}+0.75$ \GeV{}$)$,
where $m_{\Upsilon(1S)}=9.460$~\GeV and $m_{\Upsilon(3S)}=10.355$~\GeV~\cite{Zyla:2020zbs},
are vetoed to suppress the quarkonia background.

Events passing the above selections, referred to as the signal region~(SR), are used to search for the dark photon.
The dominant background contribution in the SR is from the \qqllll\ process.
The kinematic distributions of the \qqllll\ background are modeled using simulation, while the background event yield is normalized to data with the help of a control region (CR) enriched in \qqllll\ events.
The CR is defined similarly to the SR but with $m_{Z} - 5~\GeV{} < m_{4\ell} < m_{Z} + 5~\GeV{}$,
and the $\mlltwo \addslash \mllone$ and $\Upsilon$ veto requirements are not applied.
The modeling of the kinematic properties of the \qqllll\ background is studied in a validation region (VR),
which is disjoint to both the SR and the CR.
The VR is defined using the same selections as for the SR except for requiring $\mlltwo \addslash \mllone < 0.85$.
The selection efficiency of signal samples varies from 8\% to 18\% in the SR.
The fraction of the signal yield in the CR~(VR) over the SR is generally less than 5\%~(15\%) when $\mhD >2 \mdph$, but it becomes large when \mdph\ is close to \mhD, due to the presence of an off-shell \dph.

Subleading background originates from processes involving the production of \Zjet, top-quark and $WZjj$ events, with nonprompt leptons from hadron decays or misidentification of jets.
A fake-factor method described in Ref.~\cite{STDM-2016-15} is used to estimate the contributions from non-prompt leptons.
The fake factor is defined as the ratio of numbers of nonprompt leptons $N_{\mathrm{fake}}^{\mathrm{identified}}/N_{\mathrm{fake}}^{\mathrm{anti-identified}}$, where \enquote{identified} or \enquote{anti-identified} indicate whether those leptons pass all the requirements on the impact parameters, isolation and identification, or fail at least one of the requirements.
The fake factor is measured in \Zjet events, using additional leptons and not the lepton pair arising from the $Z$ boson decay.
The nonprompt-lepton background is then estimated by applying the fake factor in a region defined with the same event selection as the SR, but with at least one anti-identified lepton required when forming the quadruplet.
Minor background contributions from $pp\to H\to 4\ell$, the $gg\rightarrow ZZ\to 4\ell$ continuum, and $VVV$ and $t\bar{t}Z$ processes are estimated from simulation,
and their event yield contribution is found to be about 5\% in the SR.

The search sensitivity is limited by statistical uncertainties. Systematic uncertainties associated with the prediction of signal and background processes are also considered.
These uncertainties are either experimental or theoretical in nature,
due to imperfect modeling of the detector in the simulation or the underlying physics of each process.
Experimental uncertainties originate mainly from measurements of lepton energies, and lepton reconstruction and identification efficiencies.
Uncertainties due to the trigger selection efficiency, \pileup correction, and luminosity measurement are also considered.
Overall, the total experimental uncertainty in the predicted yields is about $7\%$ ($6\%$) for the signal (background with prompt leptons).
The theoretical uncertainties of the signal, as well as the major background due to the \qqllll\ process, include the uncertainties from PDFs, QCD scales, and \alphas.
The PDF uncertainty is estimated following the PDF4LHC~\cite{Butterworth:2015oua} procedure.
The \alphas uncertainty's effect is estimated by varying the nominal \alphas\ value of 0.118 by $\pm0.001$.
The QCD scale uncertainty's effect is estimated by varying the renormalization and factorization scales, following the procedure described in Ref.~\cite{STDM-2017-03}.
The parton showering and hadronization uncertainty is estimated for the signal by comparing the nominal \PYTHIA[8] parton showering with the alternative \HERWIG[7]~\cite{Bellm:2015jjp, Bahr:2008pv} algorithm.
For the \qqllll\ background, the modeling uncertainty due to the ME, showering, and hadronization is obtained by comparing predictions from the nominal \SHERPA sample and an alternative sample generated by \POWHEGBOX[v2] interfaced with \PYTHIA[8].
Modeling uncertainties in the $p_{\mathrm{T}}^{Z}$ distribution for the signal process, which is simulated at LO, are also considered.
The total theoretical uncertainties in the reconstructed event yields for the signal and the \qqllll\ background processes are estimated to be about 14\% and 13\%, respectively.
Systematic uncertainties assigned to the fake-lepton background,
mainly account for differences in the composition of the events with fake leptons
between \Zjet\ events and the events in the SR, and data statistical uncertainties in the dedicated region where fake factors are applied.
They are estimated to be about 51\% and 41\%, respectively, with a total uncertainty of 66\% in the fake-lepton background yield..

A simultaneous profiled binned maximum-likelihood fit~\cite{Moneta:2010pm,Verkerke:2003ir,Cranmer:1456844} to the average invariant mass \mmll, $\mmll = (\mllone + \mlltwo)/2$,
of events in the SR and CR is performed to constrain uncertainties and obtain information on a possible signal.
The experimental resolution of \mmll\ is about 1.7\% relative to \mdph\ for all simulated signal samples.
A bin width of 1~\GeV\ is used for \mmll\ distributions to take into account the resolution of the signal samples and data statistical uncertainties.
The normalizations of both the signal and the \qqllll\ background are allowed to float in the fit.
Systematic uncertainties described above are modeled as constrained nuisance parameters.
A background-only fit is also performed and the obtained background prediction is compared with data in the VR to assess the quality of the background modeling.

Table~\ref{tab:SR-yield} shows the expected background and observed event yields in the SR, CR and VR after the background-only fit.
The normalization factor of the \qqllll\ background is determined to be $0.95\pm 0.08$.
The \mmll\ distributions in the SR, CR and VR are presented in Figure~\ref{fig:kine}.
The data are found to be consistent with the background expectation in all three regions.
No significant deviation from the SM background hypothesis is observed and the largest excess of events is found around $\mmll = 25$~\GeV, with a local significance of about $1.6\sigma$. Exclusion limits are set using the CL$_\mathrm{s}$ prescription~\cite{Read:2002hq}.
Upper limits at 95\% confidence level (CL) on the cross-section times branching fraction of the process \procname\ are shown in Figure~\ref{fig:Limit_xs_LimitBand} as a function of \mdph for different \hD masses. The lower sensitivity in the mass range $\mdph > \mhD/2$ is due to the smaller signal acceptance for the off-shell \dph, and the gap around 10~\GeV\ is due to the $\Upsilon$ veto requirement in the event selection.
Since it is assumed that the SM and dark Higgs bosons do not mix, and that \dph\ is the lightest particle in the DS,
the branching fraction for \hD decay into a \dph\ pair, and for \dph\ decay into a SM fermion pair, is set to 100\%.
The branching fraction for \dph\ decay into a specific fermion pair is dependent on \mdph~\cite{Batell:2009yf,Curtin:2014cca}.
In this dark Abelian Higgs model, upper limits at 90\% CL are also set on the parameter combination $\alD\epsl^{2}$, which scales the signal yield linearly,
as shown in Figure~\ref{fig:Limit_param_compare_Belle}.
The search is sensitive to a set of \mdph and \mhD masses complementary to, and higher than, those in a similar search reported by the Belle Collaboration~\cite{Jaegle:2015fme}.
 
\begin{table}[ht]
\centering{
\caption{Post-fit expected background and observed number of data events in the SR, CR and VR.
The \enquote{Fake} background represents the contribution from non-prompt leptons, and the \enquote{Others} category combines $gg\to ZZ$, Higgs, $VVV$ and $t\bar{t}Z$ background contributions.
The expected signal yields for three benchmark points are also shown, with cross-sections calculated with $\alD=0.1$ and $\epsl=10^{-3}$.
\label{tab:SR-yield}}
\small
\begin{tabular}{ c | c | c | c }
\hline\hline
SM backgrounds & SR  & CR & VR \\
\hline
$qq \to 4\ell$ & 26.0 $\pm$ 2.4 & 1555 $\pm$ 48 & 239 $\pm$ 15 \\
\hline
Fake & 13.2 $\pm$ 5.6 & 43 $\pm$ 25 & 47 $\pm$ 26 \\
\hline
Others & 2.2 $\pm$ 0.7 & 5.8 $\pm$ 1.9 & 6.8 $\pm$ 2.0 \\
\hline
Total background & 41.3  $\pm$ 5.3 & 1604 $\pm$ 40 & 293 $\pm$ 28 \\
\hline\hline
Data  &  44 & 1602 & 286 \\
\hline\hline
Signal (\mdph, \mhD) = (12, 40)~\GeV{} & 5.0 $\pm$ 0.8 & (11.6 $\pm$ 1.4)$\times10^{-2}$ & 0.66 $\pm$ 0.10 \\
\hline
Signal (\mdph, \mhD) = (25, 40)~\GeV{} & 5.1 $\pm$ 0.8 & (11.3 $\pm$ 1.7)$\times10^{-2}$ & 0.85 $\pm$ 0.14 \\
\hline
Signal (\mdph, \mhD) = (35, 40)~\GeV{} & 6.8 $\pm$ 1.0 & 2.9 $\pm$ 0.4 & 0.67 $\pm$ 0.10 \\
\hline\hline
\end{tabular}
}
\end{table}
 
\begin{figure}[htb]
\centering
\null\hfill
\subfigure{\includegraphics[width = 0.32 \textwidth]{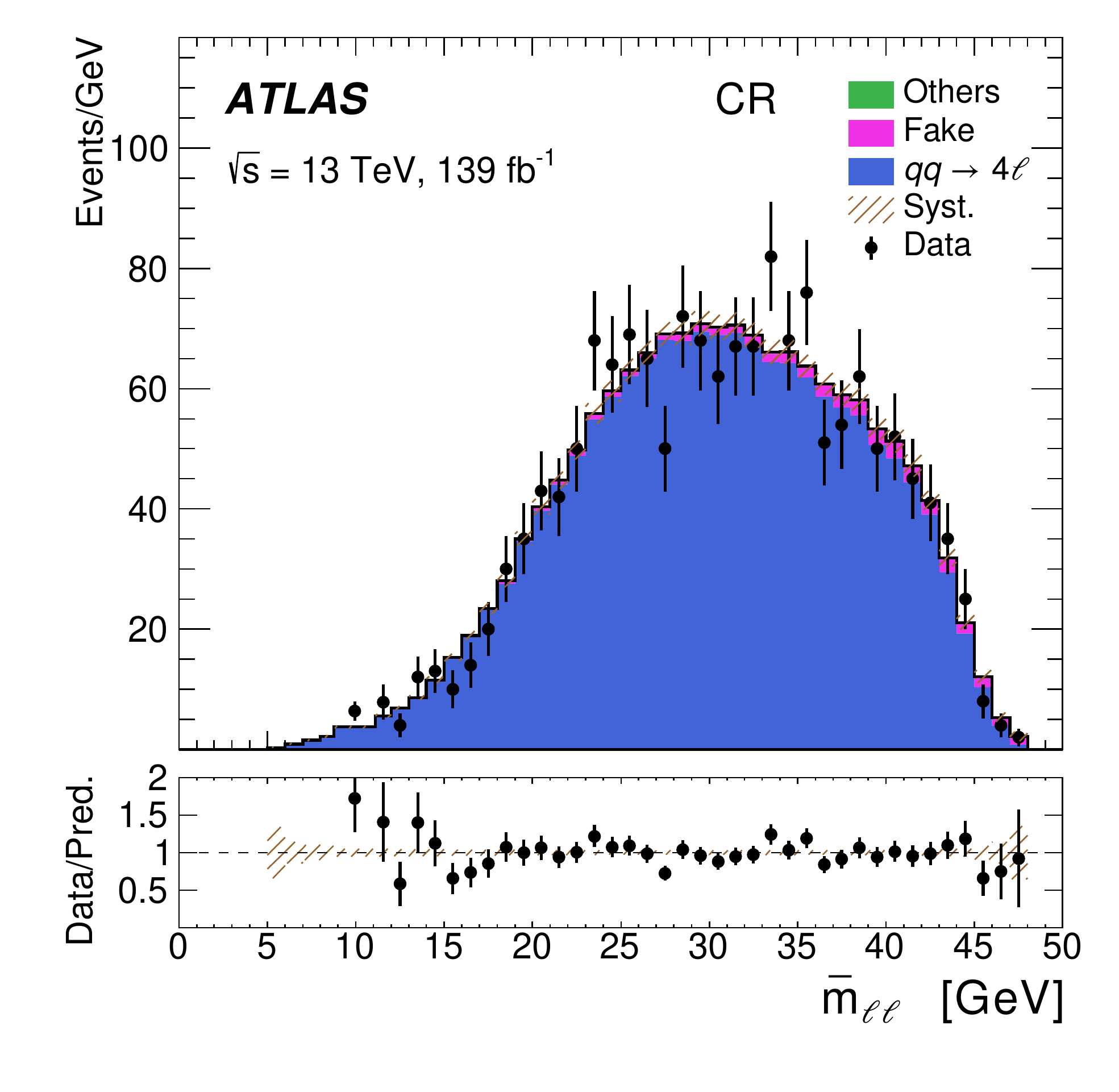}\label{fig:kine_b}}
\hfill
\subfigure{\includegraphics[width = 0.32 \textwidth]{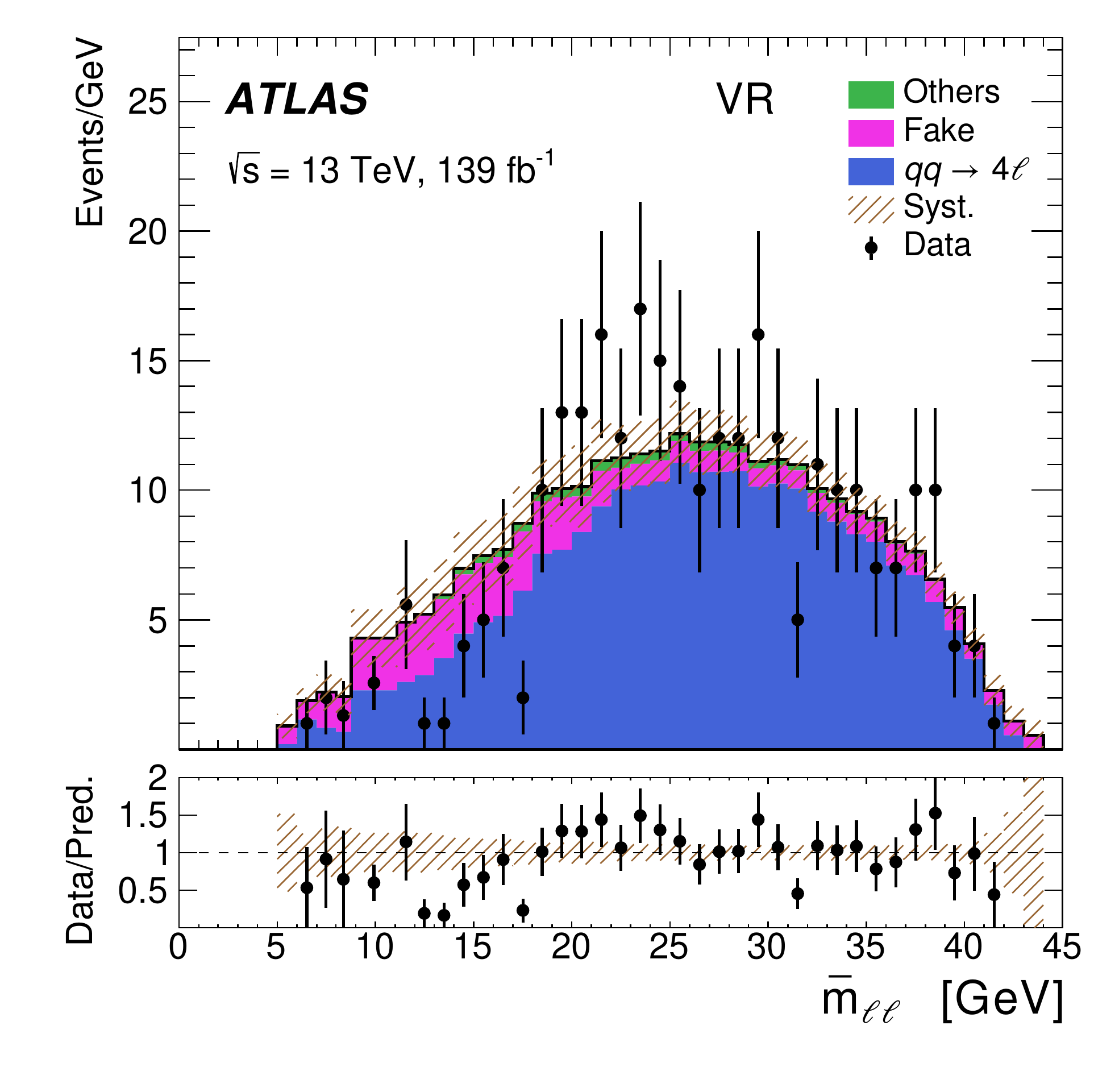}\label{fig:kine_c}}
\hfill
\subfigure{\includegraphics[width = 0.32 \textwidth]{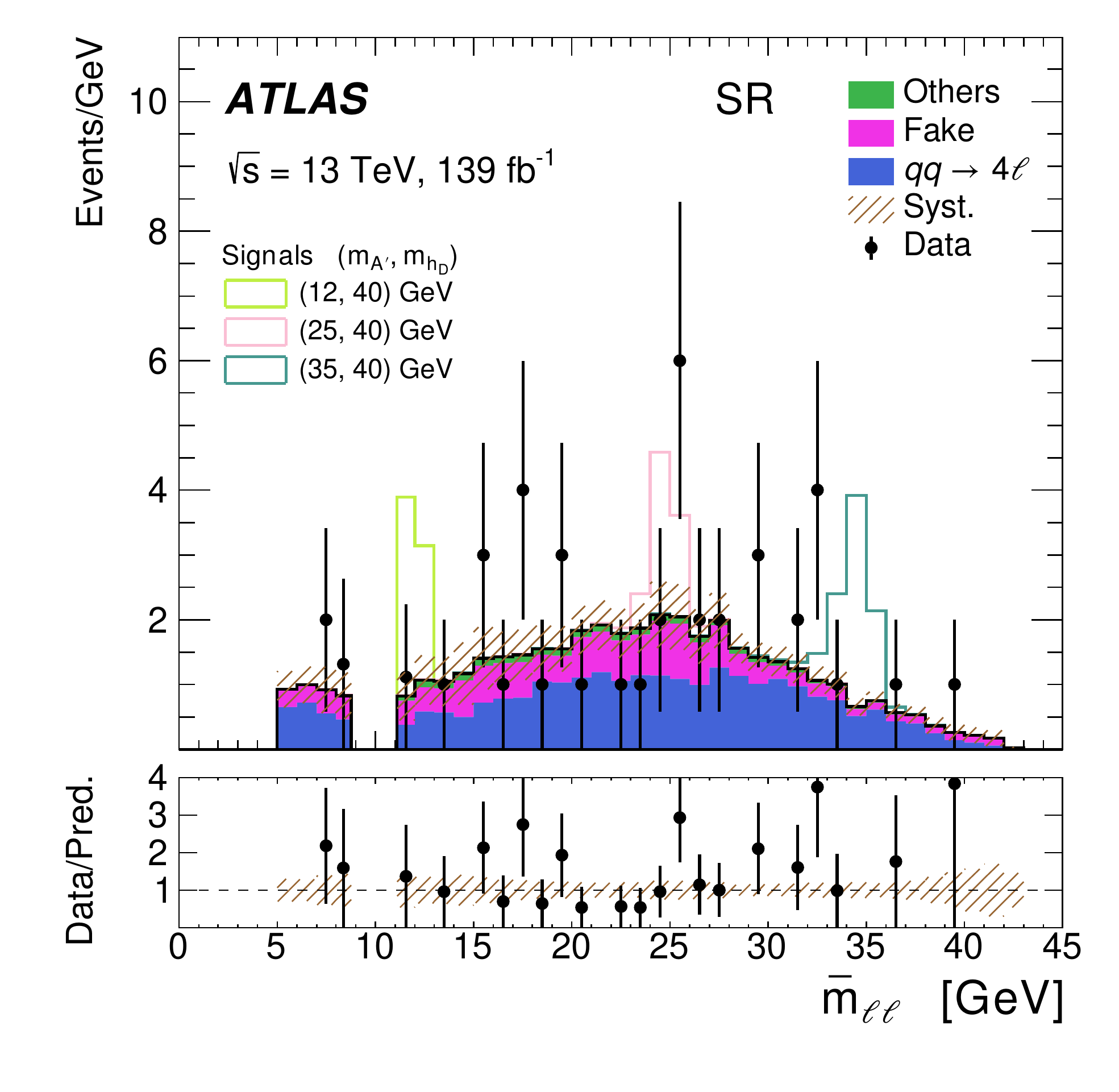}\label{fig:kine_a}}
\hfill\null
\caption{The \mmll\ distribution in the CR, VR, and SR for the data and post-fit background contributions.
The error bands include experimental and theoretical systematic uncertainties as constrained by a background-only fit.
The contributions from the production of \qqllll\ events are scaled by a normalization factor 0.95,
from the simultaneous fit in the SR and CR.
The \enquote{Others} category combines $gg\to ZZ$, Higgs, $VVV$ and $t\bar{t}Z$ background contributions.
The \enquote{Fake} background represents the contribution from non-prompt leptons.
Three representative signal distributions are overlaid in the SR, assuming $\mhD$ = 40~\GeV{} and different values of $\mdph$.
The cross-sections for these benchmark points are calculated with $\alD=0.1$ and $\epsl=10^{-3}$.
}
\label{fig:kine}
\end{figure}

\begin{figure}[htb]
\centering
\includegraphics[width = 1 \textwidth]{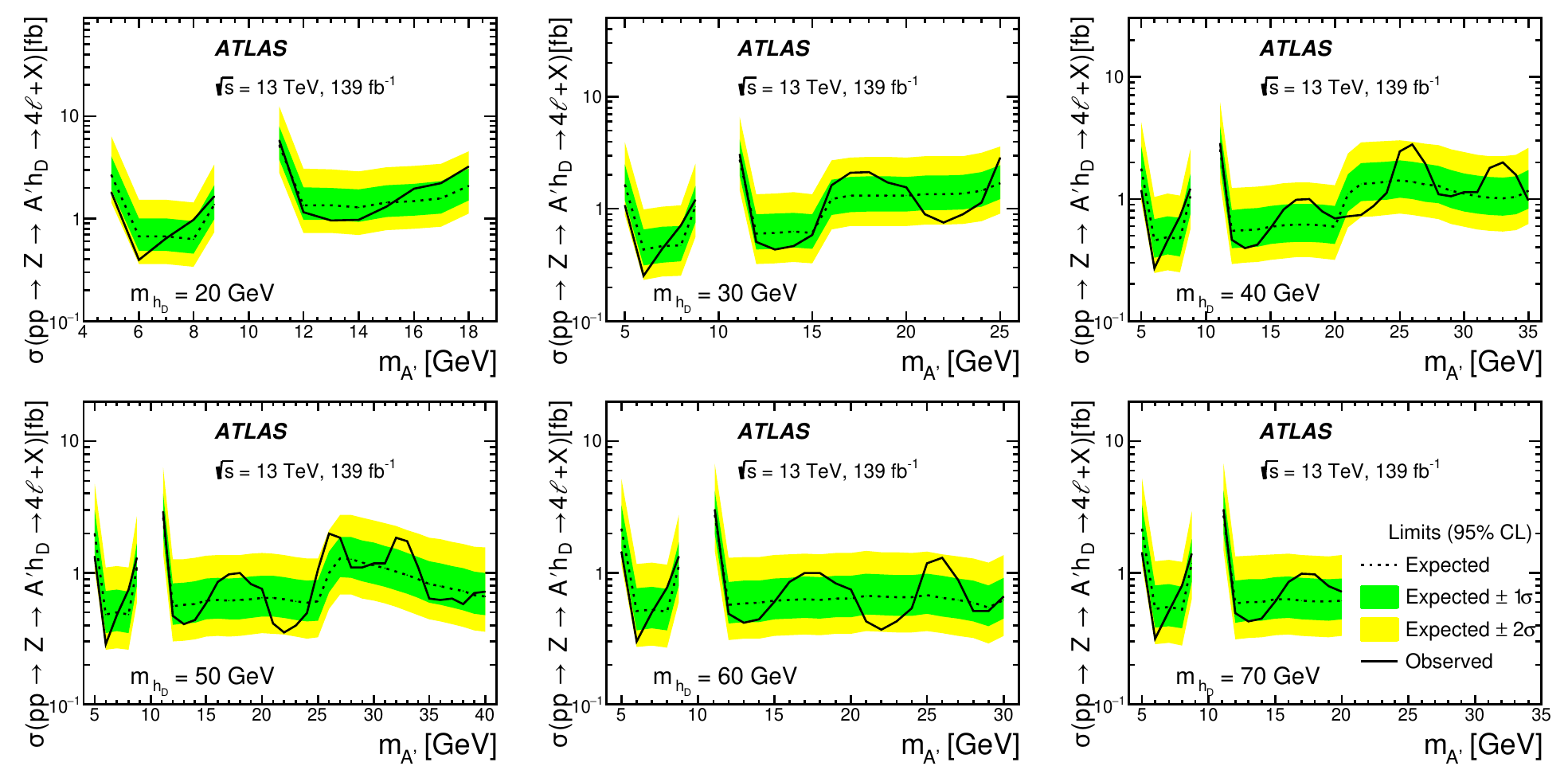}
\caption{Observed and expected upper limits at 95\% CL on the production cross-section times branching fraction as a function of $\mdph$, from top left to bottom right, corresponding to the dark Higgs boson mass of 20~GeV, 30~GeV, 40~GeV, 50~GeV, 60~GeV, and 70~GeV, respectively.
The green~(inner) and yellow~(outer) bands represent the $\pm 1\sigma$ and $\pm 2\sigma$ uncertainty in the expected limits.}
\label{fig:Limit_xs_LimitBand}
\end{figure}
 
\begin{figure}
\centering
\includegraphics[width = 0.65 \textwidth]{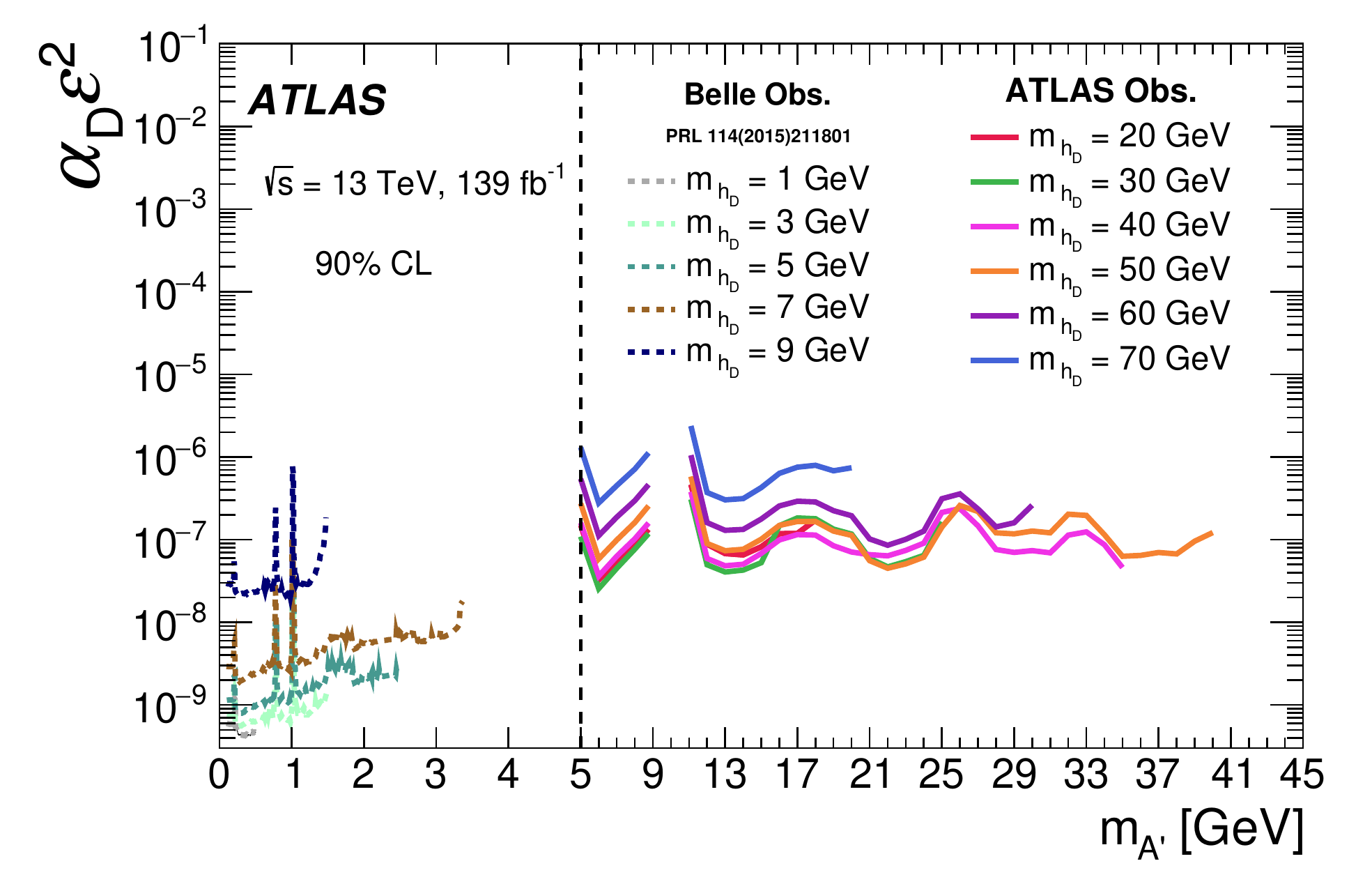}
\caption{Observed 90\% CL upper limits on $\alD\epsl^{2}$, as a function of \mdph with different dark Higgs boson masses, from this search (solid curves) compared with the results from Belle~\cite{Jaegle:2015fme}~(dashed curves).}
\label{fig:Limit_param_compare_Belle}
\end{figure}

Figure~\ref{fig:Limit_param_LimitMedian} shows the upper limits at 90\% CL on $\epsl^{2}$ as a function of \mdph with different dark Higgs boson masses,
with a benchmark value of $\alD = 0.1$ as used elsewhere~\cite{Izaguirre:2017bqb,Mohlabeng:2019vrz,NA64:2021acr,SHiP:2020noy}.
These are compared with recent results from the LHCb~\cite{LHCb:2019vmc} and CMS~\cite{CMS:2019buh} collaborations,
using the process $pp\to \dph \to \mu^{+}\mu^{-}$, which does not depend on \alD.
For $\mhD \lesssim 60$~\GeV\ and $\alD \gtrsim 0.1$, the exclusion sensitivity of this search is comparable to, or better than, that of the LHCb and CMS searches.
 
\begin{figure}
\centering
\includegraphics[width = 0.65 \textwidth]{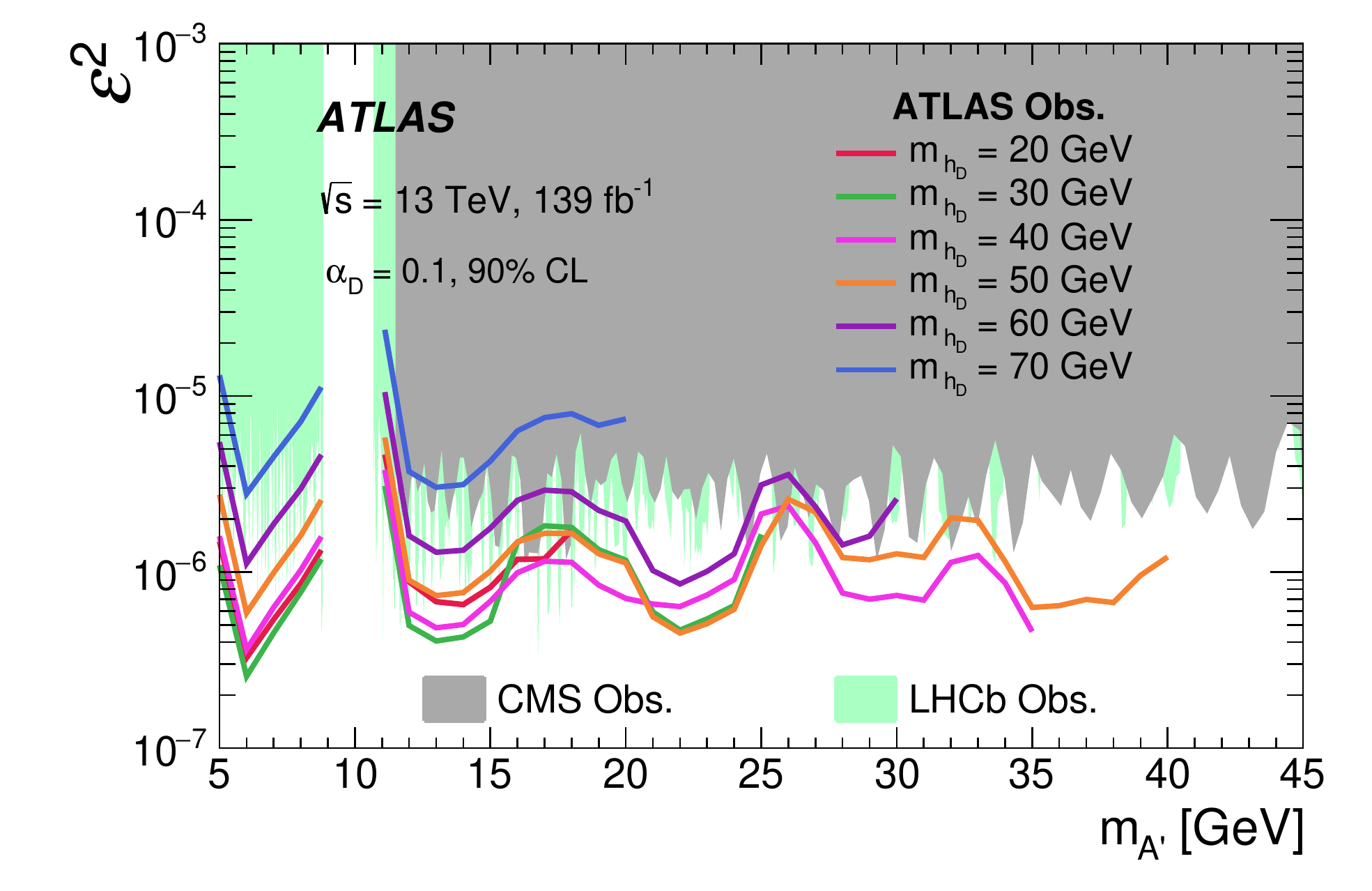}
\caption{Observed 90\% CL upper limits on $\epsl^{2}$, assuming of $\alD=0.1$, as a function of \mdph with different dark Higgs boson masses ranging from 20~ \GeV{} to 70~ \GeV{}. The parameter space excluded by LHCb~\cite{LHCb:2019vmc} (CMS~\cite{CMS:2019buh}) is covered by the green (gray) shaded regions.}
\label{fig:Limit_param_LimitMedian}
\end{figure}


In conclusion, this Letter reports the first search for a dark photon and dark Higgs boson produced via the dark Higgs-strahlung process in rare $Z$ boson decays at the LHC,
with a final state of at least four charged leptons and using 139~\ifb~of $\sqrt{s} = 13$~\TeV\ $pp$ collision data recorded by the ATLAS detector.
The data are found to be consistent with the background prediction. Upper limits are set on
the production cross-section times branching fraction, $\sigma(\procname)$,
and on the dark photon coupling to the dark Higgs boson times the kinetic mixing between the Standard Model photon and the dark photon,
$\alD\epsl^2$, in the mass ranges of $5~\GeV <\mdph < 40~\GeV$ and $20~\GeV < \mhD < 70~\GeV$.
This search explores new regions of parameter space not previously excluded by other experiments.
 
\clearpage


\section*{Acknowledgements}


We thank CERN for the very successful operation of the LHC, as well as the
support staff from our institutions without whom ATLAS could not be
operated efficiently.
 
We acknowledge the support of
ANPCyT, Argentina;
YerPhI, Armenia;
ARC, Australia;
BMWFW and FWF, Austria;
ANAS, Azerbaijan;
CNPq and FAPESP, Brazil;
NSERC, NRC and CFI, Canada;
CERN;
ANID, Chile;
CAS, MOST and NSFC, China;
Minciencias, Colombia;
MEYS CR, Czech Republic;
DNRF and DNSRC, Denmark;
IN2P3-CNRS and CEA-DRF/IRFU, France;
SRNSFG, Georgia;
BMBF, HGF and MPG, Germany;
GSRI, Greece;
RGC and Hong Kong SAR, China;
ISF and Benoziyo Center, Israel;
INFN, Italy;
MEXT and JSPS, Japan;
CNRST, Morocco;
NWO, Netherlands;
RCN, Norway;
MEiN, Poland;
FCT, Portugal;
MNE/IFA, Romania;
MESTD, Serbia;
MSSR, Slovakia;
ARRS and MIZ\v{S}, Slovenia;
DSI/NRF, South Africa;
MICINN, Spain;
SRC and Wallenberg Foundation, Sweden;
SERI, SNSF and Cantons of Bern and Geneva, Switzerland;
MOST, Taiwan;
TENMAK, T\"urkiye;
STFC, United Kingdom;
DOE and NSF, United States of America.
In addition, individual groups and members have received support from
BCKDF, CANARIE, Compute Canada and CRC, Canada;
PRIMUS 21/SCI/017 and UNCE SCI/013, Czech Republic;
COST, ERC, ERDF, Horizon 2020, ICSC-NextGenerationEU and Marie Sk{\l}odowska-Curie Actions, European Union;
Investissements d'Avenir Labex, Investissements d'Avenir Idex and ANR, France;
DFG and AvH Foundation, Germany;
Herakleitos, Thales and Aristeia programmes co-financed by EU-ESF and the Greek NSRF, Greece;
BSF-NSF and MINERVA, Israel;
Norwegian Financial Mechanism 2014-2021, Norway;
NCN and NAWA, Poland;
La Caixa Banking Foundation, CERCA Programme Generalitat de Catalunya and PROMETEO and GenT Programmes Generalitat Valenciana, Spain;
G\"{o}ran Gustafssons Stiftelse, Sweden;
The Royal Society and Leverhulme Trust, United Kingdom.
 
The crucial computing support from all WLCG partners is acknowledged gratefully, in particular from CERN, the ATLAS Tier-1 facilities at TRIUMF (Canada), NDGF (Denmark, Norway, Sweden), CC-IN2P3 (France), KIT/GridKA (Germany), INFN-CNAF (Italy), NL-T1 (Netherlands), PIC (Spain), ASGC (Taiwan), RAL (UK) and BNL (USA), the Tier-2 facilities worldwide and large non-WLCG resource providers. Major contributors of computing resources are listed in Ref.~\cite{ATL-SOFT-PUB-2023-001}.


\clearpage
\appendix

\clearpage
\part*{Appendix}
 
Upper limits at 95\% CL are also set on the branching fraction of the rare $Z$ boson decay $Z\rightarrow \dph\hD$,
\begin{align*}
\brZ =\frac{\xsproc)}{\xsprocZ \cdot \brdp},
\end{align*}
where \xsproc\ is the 95\% CL upper limit taken from Figure~\ref{fig:Limit_xs_LimitBand},
\brdp\ is the branching fraction of \dph and \hD decaying into at least two lepton pairs,
and \xsprocZ\ is the measured $Z$ boson production cross-section as described in Ref.~\cite{STDM-2016-02} in the phase space of $66~\GeV < \mll < 116~\GeV$.
The branching fraction limits are shown in Figure~\ref{fig:Limit_br_LimitBand}.
 
\begin{figure}[htb]
\centering
\includegraphics[width = 0.9 \textwidth]{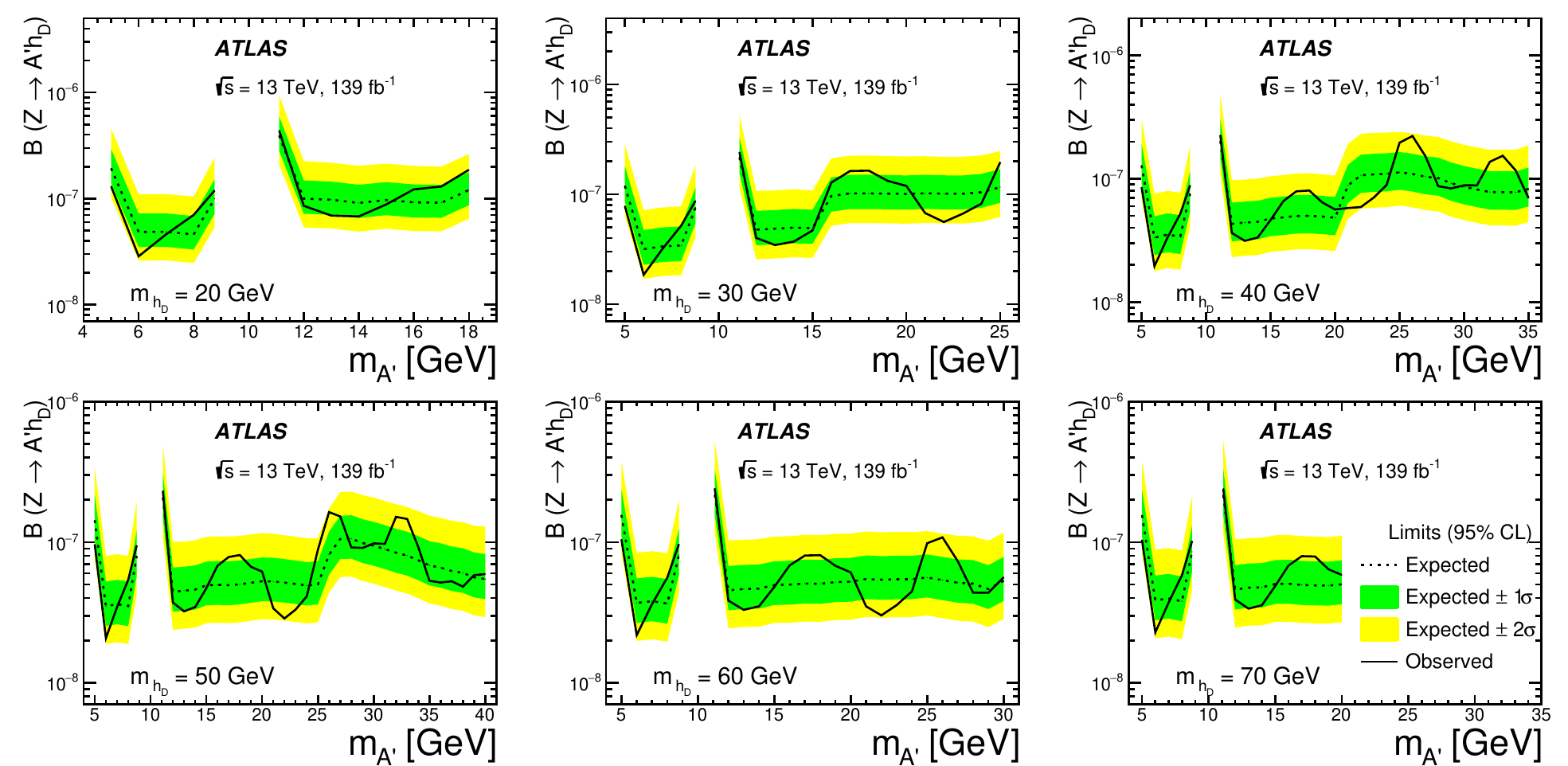}
\caption{Observed and expected upper limits at 95\% CL on the branching fraction \brZ\ as a function of $\mdph$, from top left to bottom right corresponding to the dark Higgs boson mass of 20~\GeV, 30~\GeV, 40~\GeV, 50~\GeV, 60~\GeV, and 70~\GeV, respectively.
The green~(inner) and yellow~(outer) bands represent the $\pm 1\sigma$ and $\pm 2\sigma$ uncertainty in the expected limits.}
\label{fig:Limit_br_LimitBand}
\end{figure}

\clearpage

\printbibliography

@book{Bertone:2010zza,
    author = "Silk, J. and others",
    editor = "Bertone, Gianfranco",
    title = "{Particle Dark Matter: Observations, Models and Searches}",
    doi = "10.1017/CBO9780511770739",
    isbn = "978-1-107-65392-4",
    publisher = "Cambridge Univ. Press",
    address = "Cambridge",
    year = "2010"
}

@article{Feng:2010gw,
    author = "Feng, Jonathan L.",
    title = "{Dark Matter Candidates from Particle Physics and Methods of Detection}",
    eprint = "1003.0904",
    archivePrefix = "arXiv",
    primaryClass = "astro-ph.CO",
    reportNumber = "UCI-TR-2009-13",
    doi = "10.1146/annurev-astro-082708-101659",
    journal = "Ann. Rev. Astron. Astrophys.",
    volume = "48",
    pages = "495--545",
    year = "2010"
}

@article{Porter:2011nv,
    author = "Porter, Troy A. and Johnson, Robert P. and Graham, Peter W.",
    title = "{Dark Matter Searches with Astroparticle Data}",
    eprint = "1104.2836",
    archivePrefix = "arXiv",
    primaryClass = "astro-ph.HE",
    doi = "10.1146/annurev-astro-081710-102528",
    journal = "Ann. Rev. Astron. Astrophys.",
    volume = "49",
    pages = "155--194",
    year = "2011"
}

@article{Bertone:2017adx,
    author = "Bertone, Gianfranco and Bozorgnia, Nassim and Kim, Jong Soo and Liem, Sebastian and McCabe, Christopher and Otten, Sydney and Ruiz de Austri, Roberto",
    title = "{Identifying WIMP dark matter from particle and astroparticle data}",
    eprint = "1712.04793",
    archivePrefix = "arXiv",
    primaryClass = "hep-ph",
    reportNumber = "IPPP-17-94, KCL-PH-TH-2017-55, TTK-17-45",
    doi = "10.1088/1475-7516/2018/03/026",
    journal = "JCAP",
    volume = "03",
    pages = "026",
    year = "2018"
}

@article{Holdom:1985ag,
    author = "Holdom, Bob",
    title = "{Two U(1)'s and $\epsilon$ charge shifts}",
    reportNumber = "UTPT-85-30",
    doi = "10.1016/0370-2693(86)91377-8",
    journal = "Phys. Lett. B",
    volume = "166",
    pages = "196--198",
    year = "1986"
}

@article{Galison:1983pa,
    author = "Galison, Peter and Manohar, Aneesh",
    title = "{Two Z's or not two Z's?}",
    reportNumber = "HUTP-83/A059",
    doi = "10.1016/0370-2693(84)91161-4",
    journal = "Phys. Lett. B",
    volume = "136",
    pages = "279--283",
    year = "1984"
}

@article{Batell:2009yf,
    author = "Batell, Brian and Pospelov, Maxim and Ritz, Adam",
    title = "{Probing a Secluded U(1) at B-factories}",
    eprint = "0903.0363",
    archivePrefix = "arXiv",
    primaryClass = "hep-ph",
    doi = "10.1103/PhysRevD.79.115008",
    journal = "Phys. Rev. D",
    volume = "79",
    pages = "115008",
    year = "2009"
}

@article{Bjorken:2009mm,
    author = "Bjorken, James D. and Essig, Rouven and Schuster, Philip and Toro, Natalia",
    title = "{New Fixed-Target Experiments to Search for Dark Gauge Forces}",
    eprint = "0906.0580",
    archivePrefix = "arXiv",
    primaryClass = "hep-ph",
    reportNumber = "SLAC-PUB-13650, SU-ITP-09-22",
    doi = "10.1103/PhysRevD.80.075018",
    journal = "Phys. Rev. D",
    volume = "80",
    pages = "075018",
    year = "2009"
}

@article{Marsicano:2018krp,
    author = "Marsicano, L. and Battaglieri, M. and Bondi', M. and Carvajal, C. D. R. and Celentano, A. and De~Napoli, M. and De~Vita, R. and Nardi, E. and Raggi, M. and Valente, P.",
    title = "{Dark photon production through positron annihilation in beam-dump experiments}",
    eprint = "1802.03794",
    archivePrefix = "arXiv",
    primaryClass = "hep-ex",
    doi = "10.1103/PhysRevD.98.015031",
    journal = "Phys. Rev. D",
    volume = "98",
    number = "1",
    pages = "015031",
    year = "2018"
}

@article{Jaegle:2015fme,
    author = "{Belle Collaboration}",
    collaboration = "Belle",
    title = "{Search for the dark photon and the dark Higgs boson at Belle}",
    eprint = "1502.00084",
    archivePrefix = "arXiv",
    primaryClass = "hep-ex",
    reportNumber = "Belle-Preprint-2014-20, KEK-Preprint-2014-36",
    doi = "10.1103/PhysRevLett.114.211801",
    journal = "Phys. Rev. Lett.",
    volume = "114",
    number = "21",
    pages = "211801",
    year = "2015"
}

@article{Belle-II:2022jyy,
    author = "{Belle-II Collaboration}",
    collaboration = "Belle-II",
    title = "{Search for a Dark Photon and an Invisible Dark Higgs Boson in $\ensuremath{\mu}^{+}\ensuremath{\mu}^{-}$ and Missing Energy Final States with the Belle II Experiment}",
    eprint = "2207.00509",
    archivePrefix = "arXiv",
    primaryClass = "hep-ex",
    reportNumber = "Belle II Preprint 2022-001, KEK Preprint 2022-6",
    doi = "10.1103/PhysRevLett.130.071804",
    journal = "Phys. Rev. Lett.",
    volume = "130",
    number = "7",
    pages = "071804",
    year = "2023"
}

@article{KLOE-2:2015nli,
    author = "{KLOE-2 Collaboration}",
    collaboration = "KLOE-2",
    title = "{Search for dark Higgsstrahlung in $e^{+}e^{-}\to\ensuremath{\mu}^{+}\ensuremath{\mu}^{-}$ and missing energy events with the KLOE experiment}",
    eprint = "1501.06795",
    archivePrefix = "arXiv",
    primaryClass = "hep-ex",
    doi = "10.1016/j.physletb.2015.06.015",
    journal = "Phys. Lett. B",
    volume = "747",
    pages = "365--372",
    year = "2015"
}

@article{BaBar:2012bkw,
    author = "{BABAR Collaboration}",
    collaboration = "BaBar",
    title = "{Search for Low-Mass Dark-Sector Higgs Bosons}",
    eprint = "1202.1313",
    archivePrefix = "arXiv",
    primaryClass = "hep-ex",
    reportNumber = "BABAR-PUB-11-025, SLAC-PUB-14863",
    doi = "10.1103/PhysRevLett.108.211801",
    journal = "Phys. Rev. Lett.",
    volume = "108",
    pages = "211801",
    year = "2012"
}

@article{KLOE-2:2016ydq,
    author = "{KLOE-2 Collaboration}",
    collaboration = "KLOE-2",
    title = "{Limit on the production of a new vector boson in $\mathrm{e^+ e^-}\rightarrow U\gamma$, U$\rightarrow \pi^+\pi^-$ with the KLOE experiment}",
    eprint = "1603.06086",
    archivePrefix = "arXiv",
    primaryClass = "hep-ex",
    doi = "10.1016/j.physletb.2016.04.019",
    journal = "Phys. Lett. B",
    volume = "757",
    pages = "356--361",
    year = "2016"
}

@article{Blinov:2017dtk,
    author = "Blinov, Nikita and Izaguirre, Eder and Shuve, Brian",
    title = "{Rare $Z$ Boson Decays to a Hidden Sector}",
    eprint = "1710.07635",
    archivePrefix = "arXiv",
    primaryClass = "hep-ph",
    reportNumber = "SLAC-PUB-17159",
    doi = "10.1103/PhysRevD.97.015009",
    journal = "Phys. Rev. D",
    volume = "97",
    number = "1",
    pages = "015009",
    year = "2018"
}

@article{Curtin:2014cca,
    author = "Curtin, David and Essig, Rouven and Gori, Stefania and Shelton, Jessie",
    title = "{Illuminating Dark Photons with High-Energy Colliders}",
    eprint = "1412.0018",
    archivePrefix = "arXiv",
    primaryClass = "hep-ph",
    reportNumber = "YITP-SB-14-49",
    doi = "10.1007/JHEP02(2015)157",
    journal = "JHEP",
    volume = "02",
    pages = "157",
    year = "2015"
}

@article{Lai_2010,
      author         = "Lai, H. L. and Guzzi, M. and Huston, J. and Li, Z. and Nadolsky, P. M. and Pumplin, J. and Yuan, C. P.",
      title          = "{New parton distributions for collider physics}",
      journal        = "Phys. Rev. D",
      volume         = "82",
      pages          = "074024",
      doi            = "10.1103/PhysRevD.82.074024",
      year           = "2010",
      eprint         = "1007.2241",
      archivePrefix  = "arXiv",
      primaryClass   = "hep-ph",
      reportNumber   = "MSUHEP-100707, SMU-HEP-10-10",
      SLACcitation   = "%%CITATION = ARXIV:1007.2241;%%",
}

@article{Caola_2015,
    author = {Caola, Fabrizio and Melnikov, Kirill and R{\"o}ntsch, Raoul and Tancredi, Lorenzo},
    title = "{QCD corrections to ZZ production in gluon fusion at the LHC}",
    eprint = "1509.06734",
    archivePrefix = "arXiv",
    primaryClass = "hep-ph",
    reportNumber = "TTP15-035, CERN-PH-TH-2015-226, FERMILAB-PUB-15-398-T",
    doi = "10.1103/PhysRevD.92.094028",
    journal = "Phys. Rev. D",
    volume = "92",
    number = "9",
    pages = "094028",
    year = "2015"
}

@article{Hamilton_2013,
    author = "Hamilton, Keith and Nason, Paolo and Re, Emanuele and Zanderighi, Giulia",
    title = "{NNLOPS simulation of Higgs boson production}",
    eprint = "1309.0017",
    archivePrefix = "arXiv",
    primaryClass = "hep-ph",
    reportNumber = "MCNET-13-11, CERN-PH-TH-2013-205, OUTP-13-18P",
    doi = "10.1007/JHEP10(2013)222",
    journal = "JHEP",
    volume = "10",
    pages = "222",
    year = "2013"
}

@article{Nason_2010,
    author = "Nason, Paolo and Oleari, Carlo",
    title = "{NLO Higgs boson production via vector-boson fusion matched with shower in POWHEG}",
    eprint = "0911.5299",
    archivePrefix = "arXiv",
    primaryClass = "hep-ph",
    doi = "10.1007/JHEP02(2010)037",
    journal = "JHEP",
    volume = "02",
    pages = "037",
    year = "2010"
}

@article{Verkerke:2003ir,
    author = "Verkerke, Wouter and Kirkby, David P.",
    title = "{The RooFit toolkit for data modeling}",
    eprint = "physics/0306116",
    archivePrefix = "arXiv",
    reportNumber = "CHEP-2003-MOLT007",
    journal = "eConf",
    volume = "C0303241",
    pages = "MOLT007",
    year = "2003"
}

@article{Moneta:2010pm,
    author = "Moneta, Lorenzo and Belasco, Kevin and Cranmer, Kyle S. and Kreiss, S. and Lazzaro, Alfio and Piparo, Danilo and Schott, Gregory and Verkerke, Wouter and Wolf, Matthias",
    title = "{The RooStats Project}",
    eprint = "1009.1003",
    archivePrefix = "arXiv",
    primaryClass = "physics.data-an",
    doi = "10.22323/1.093.0057",
    journal = "PoS",
    volume = "ACAT2010",
    pages = "057",
    year = "2010"
}

@techreport{Cranmer:1456844,
      author        = "Cranmer, Kyle and Lewis, George and Moneta, Lorenzo and
                       Shibata, Akira and Verkerke, Wouter",
      title         = "{HistFactory: A tool for creating statistical models for
                       use with RooFit and RooStats}",
      institution   = "New York U.",
      collaboration = "ROOT Collaboration",
      address       = "New York",
      reportNumber  = "CERN-OPEN-2012-016",
      month         = "1",
      year          = "2012",
      url           = "https://cds.cern.ch/record/1456844",
}

@article{CMS:2019buh,
    author = "{CMS Collaboration}",
    collaboration = "CMS",
    title = "{Search for a Narrow Resonance Lighter than 200 GeV Decaying to a Pair of Muons in Proton-Proton Collisions at $\sqrt{s} =$  TeV}",
    eprint = "1912.04776",
    archivePrefix = "arXiv",
    primaryClass = "hep-ex",
    reportNumber = "CMS-EXO-19-018, CERN-EP-2019-265",
    doi = "10.1103/PhysRevLett.124.131802",
    journal = "Phys. Rev. Lett.",
    volume = "124",
    number = "13",
    pages = "131802",
    year = "2020"
}

@article{LHCb:2019vmc,
    author = "{LHCb Collaboration}",
    collaboration = "LHCb",
    title = "{Search for $A'\to\mu^+\mu^-$ Decays}",
    eprint = "1910.06926",
    archivePrefix = "arXiv",
    primaryClass = "hep-ex",
    reportNumber = "LHCb-PAPER-2019-031, CERN-EP-2019-212",
    doi = "10.1103/PhysRevLett.124.041801",
    journal = "Phys. Rev. Lett.",
    volume = "124",
    number = "4",
    pages = "041801",
    year = "2020"
}

@article{ATLAS:2022hro,
    author = "{ATLAS Collaboration}",
    collaboration = "ATLAS",
    title = "{Luminosity determination in $pp$ collisions at $\sqrt{s}=13$ TeV using the ATLAS detector at the LHC}",
    eprint = "2212.09379",
    archivePrefix = "arXiv",
    primaryClass = "hep-ex",
    reportNumber = "CERN-EP-2022-281",
    month = "12",
    year = "2022"
}

@article{Izaguirre:2017bqb,
    author = "Izaguirre, Eder and Kahn, Yonatan and Krnjaic, Gordan and Moschella, Matthew",
    title = "{Testing light dark matter coannihilation with fixed-target experiments}",
    eprint = "1703.06881",
    archivePrefix = "arXiv",
    primaryClass = "hep-ph",
    reportNumber = "FERMILAB-PUB-17-068-PPD, PUPT-2520",
    doi = "10.1103/PhysRevD.96.055007",
    journal = "Phys. Rev. D",
    volume = "96",
    number = "5",
    pages = "055007",
    year = "2017"
}

@article{Mohlabeng:2019vrz,
    author = "Mohlabeng, Gopolang",
    title = "{Revisiting the dark photon explanation of the muon anomalous magnetic moment}",
    eprint = "1902.05075",
    archivePrefix = "arXiv",
    primaryClass = "hep-ph",
    doi = "10.1103/PhysRevD.99.115001",
    journal = "Phys. Rev. D",
    volume = "99",
    number = "11",
    pages = "115001",
    year = "2019"
}

@article{NA64:2021acr,
    author = "Cazzaniga, C. and others",
    collaboration = "NA64",
    title = "{Probing the explanation of the muon (g-2) anomaly and thermal light dark matter with the semi-visible dark photon channel}",
    eprint = "2107.02021",
    archivePrefix = "arXiv",
    primaryClass = "hep-ex",
    doi = "10.1140/epjc/s10052-021-09705-5",
    journal = "Eur. Phys. J. C",
    volume = "81",
    number = "10",
    pages = "959",
    year = "2021"
}

@article{SHiP:2020noy,
    author = "Ahdida, C. and others",
    collaboration = "SHiP",
    title = "{Sensitivity of the SHiP experiment to light dark matter}",
    eprint = "2010.11057",
    archivePrefix = "arXiv",
    primaryClass = "hep-ex",
    doi = "10.1007/JHEP04(2021)199",
    journal = "JHEP",
    volume = "04",
    pages = "199",
    year = "2021"
}
 
\clearpage
 
\begin{flushleft}
\hypersetup{urlcolor=black}
{\Large The ATLAS Collaboration}

\bigskip

\AtlasOrcid[0000-0002-6665-4934]{G.~Aad}$^\textrm{\scriptsize 102}$,
\AtlasOrcid[0000-0002-5888-2734]{B.~Abbott}$^\textrm{\scriptsize 120}$,
\AtlasOrcid[0000-0002-1002-1652]{K.~Abeling}$^\textrm{\scriptsize 55}$,
\AtlasOrcid[0000-0001-5763-2760]{N.J.~Abicht}$^\textrm{\scriptsize 49}$,
\AtlasOrcid[0000-0002-8496-9294]{S.H.~Abidi}$^\textrm{\scriptsize 29}$,
\AtlasOrcid[0000-0002-9987-2292]{A.~Aboulhorma}$^\textrm{\scriptsize 35e}$,
\AtlasOrcid[0000-0001-5329-6640]{H.~Abramowicz}$^\textrm{\scriptsize 151}$,
\AtlasOrcid[0000-0002-1599-2896]{H.~Abreu}$^\textrm{\scriptsize 150}$,
\AtlasOrcid[0000-0003-0403-3697]{Y.~Abulaiti}$^\textrm{\scriptsize 117}$,
\AtlasOrcid[0000-0003-0762-7204]{A.C.~Abusleme~Hoffman}$^\textrm{\scriptsize 137a}$
% ...
\bigskip
\\

% ...
$^{29}$Physics Department, Brookhaven National Laboratory, Upton NY; United States of America.\\
% ...
$^{35}$$^{(a)}$Facult\'e des Sciences Ain Chock, R\'eseau Universitaire de Physique des Hautes Energies - Universit\'e Hassan II, Casablanca;$^{(b)}$Facult\'{e} des Sciences, Universit\'{e} Ibn-Tofail, K\'{e}nitra;$^{(c)}$Facult\'e des Sciences Semlalia, Universit\'e Cadi Ayyad, LPHEA-Marrakech;$^{(d)}$LPMR, Facult\'e des Sciences, Universit\'e Mohamed Premier, Oujda;$^{(e)}$Facult\'e des sciences, Universit\'e Mohammed V, Rabat;$^{(f)}$Institute of Applied Physics, Mohammed VI Polytechnic University, Ben Guerir; Morocco.\\
% ...
$^{49}$Fakult\"{a}t Physik , Technische Universit{\"a}t Dortmund, Dortmund; Germany.\\
% ...
$^{55}$II. Physikalisches Institut, Georg-August-Universit\"{a}t G\"ottingen, G\"ottingen; Germany.\\
% ...
$^{102}$CPPM, Aix-Marseille Universit\'e, CNRS/IN2P3, Marseille; France.\\
% ...
$^{117}$Department of Physics, New York University, New York NY; United States of America.\\
% ...
$^{120}$Homer L. Dodge Department of Physics and Astronomy, University of Oklahoma, Norman OK; United States of America.\\
% ...
$^{137}$$^{(a)}$Departamento de F\'isica, Pontificia Universidad Cat\'olica de Chile, Santiago;$^{(b)}$Millennium Institute for Subatomic physics at high energy frontier (SAPHIR), Santiago;$^{(c)}$Instituto de Investigaci\'on Multidisciplinario en Ciencia y Tecnolog\'ia, y Departamento de F\'isica, Universidad de La Serena;$^{(d)}$Universidad Andres Bello, Department of Physics, Santiago;$^{(e)}$Instituto de Alta Investigaci\'on, Universidad de Tarapac\'a, Arica;$^{(f)}$Departamento de F\'isica, Universidad T\'ecnica Federico Santa Mar\'ia, Valpara\'iso; Chile.\\
% ...
$^{150}$Department of Physics, Technion, Israel Institute of Technology, Haifa; Israel.\\
$^{151}$Raymond and Beverly Sackler School of Physics and Astronomy, Tel Aviv University, Tel Aviv; Israel.\\
% ...
$^{a}$ Also Affiliated with an institute covered by a cooperation agreement with CERN.\\
$^{b}$ Also at An-Najah National University, Nablus; Palestine.\\
$^{c}$ Also at Borough of Manhattan Community College, City University of New York, New York NY; United States of America.\\
$^{d}$ Also at Center for High Energy Physics, Peking University; China.\\
$^{e}$ Also at Center for Interdisciplinary Research and Innovation (CIRI-AUTH), Thessaloniki; Greece.\\
$^{f}$ Also at Centro Studi e Ricerche Enrico Fermi; Italy.\\
% ...

\end{flushleft}


\end{document}